\documentclass[amsmath,amssymb,aps,twocolumn,showpacs,showkeys,prl,floatfix]{revtex4-1}
\usepackage{mathrsfs}
\usepackage{amsmath}
\usepackage{amssymb}
\usepackage{color}
\usepackage{graphicx}	
\usepackage{bm}
\usepackage{ulem} 

\newcommand{\be}{\begin{equation}}
\newcommand{\ee}{\end{equation}}
\newcommand{\bef}{\begin{figure}}
\newcommand{\eef}{\end{figure}}
\newcommand{\bea}{\begin{eqnarray}}
\newcommand{\eea}{\end{eqnarray}}

\begin{document}
\title{Buckling in Armored Droplets}
\author{Fran\c cois Sicard}
\thanks{Corresponding author: \texttt{francois.sicard@free.fr}.}
\author{Alberto Striolo}
\affiliation{Department of Chemical Engineering, University College London, Torrington Place, London WC1E 7JE, United Kingdom, EU}

%
    
\begin{abstract}
The issue of the buckling mechanism in droplets stabilized by solid particles (armored droplets) 
is tackled at a mesoscopic level 
using dissipative particle dynamics simulations. We consider spherical water droplet in a decane solvent 
coated with nanoparticle monolayers of two different types: Janus and homogeneous.
The chosen particles yield comparable initial three-phase contact angles, chosen to maximize the  
adsorption energy at the interface. We study the interplay between the evolution of droplet shape, layering of 
the particles, and their distribution at the interface when the volume of the droplets is reduced. 
We show that Janus particles affect strongly the shape of the droplet with the formation of 
a crater-like depression. This evolution is \textit{actively} controlled by a close-packed particle 
monolayer at the curved interface. On the contrary, homogeneous particles follow \textit{passively} 
the volume reduction of the droplet, whose shape does not deviate too much from spherical, 
even when a nanoparticle monolayer/bilayer transition is detected at the interface.
We discuss how these buckled armored droplets might be of relevance in various applications including 
potential drug delivery systems and biomimetic design of functional surfaces.
\end{abstract}


\maketitle

Pickering emulsions~\cite{Pickering1907}, \textit{i.e.} particle-stabilized emulsions, have been studied intensively 
in recent years owning to their wide range of applications including biofuel processing~\cite{Drexler2012} 
and food preservation~\cite{Shchukina2012,Puglia2012}. They have also been developed as precursors to 
magnetic particles for imaging~\cite{LeeAM2015} and drug delivery systems~\cite{FrelichowskaIJP2009}. 
Even with their widespread use, they remain, however, underutilized. 
In Pickering emulsions, particles and/or nanoparticles (NPs) with suitable surface chemistries adsorb at the droplet surfaces, 
with an adsorption energy of up to thousands of times the thermal energy.
The characteristics of Pickering emulsions pose a number of intriguing fundamental physical questions including 
a thorough understanding of the perennial lack of detail about how particles arrange at 
the liquid/liquid interface. Other not completely answered questions include particle effects on 
interfacial tension~\cite{Miller2006}, layering~\cite{RazaviLangmuir2015}, 
buckling~\cite{TsapisPRL2005, XuLangmuir2005, DattaLangmuir2010} and 
particle release~\cite{RazaviLangmuir2015, GarbinLangmuir2011}.

In some important processes that involve emulsions, it can be required to reduce the volume of 
the dispersed droplets~\cite{PauchardPRE1999,TsapisPRL2005,GorandLangmuir2004,PauchardEL2003}.
The interface may undergo large deformations that produce compressive stresses, causing localized mechanical instabilities. The proliferation of these localized instabilities may then result in a variety of collapse 
mechanisms~\cite{XuLangmuir2005,RazaviLangmuir2015,DattaLangmuir2010}. 
Despite the vast interest in particle-laden interfaces, the key factors that determine 
the collapse of curved particle-laden interfaces are still subject of debate. 
Indeed, although linear elasticity describes successfully the morphology of buckled particle-laden droplets, 
it is still unclear whether the onset of buckling can be explained in terms of classic elastic buckling criteria~\cite{LandauBook1986,QuillietEPJE2012}, capillary pressure-driven phase transition~\cite{TsapisPRL2005}, 
or interfacial compression phase transition~\cite{SalmonLangmuir2016}.
Numerous experiments have been conducted to link the rheological response of particle-laden 
interfaces to the stability of emulsions and foams. However, their results could be dependent on the method chosen for 
preparing the interfacial layer. Due to their inherent limited resolution, \textit{direct} access to local 
observables, such as the particles three-phase contact angle distribution, remains out of reach~\cite{BinksSM2016}. 
This crucial information can be accessed by numerical simulations sometimes with approximations.
All-atom molecular dynamics (MD) simulations have become a widely employed computational technique. 
However, all-atom MD simulations are computationally expensive. 
Moreover, most phenomena of interest here take place on time scales that are orders of magnitude 
longer than those accessible via all-atom MD. Mesoscopic simulations, in which 
the structural unit is a coarse-grained representation of a large number of molecules, 
allow us to overcome these limitations. It is now well established 
that  coarse-grained approaches offer the possibility of answering fundamental questions responsible 
for the collective behaviour of particles anchored at an interface~\cite{ClarkLangmuir2000}.

\begin{figure*}[t]
\includegraphics[width=0.85 \textwidth, angle=-0]{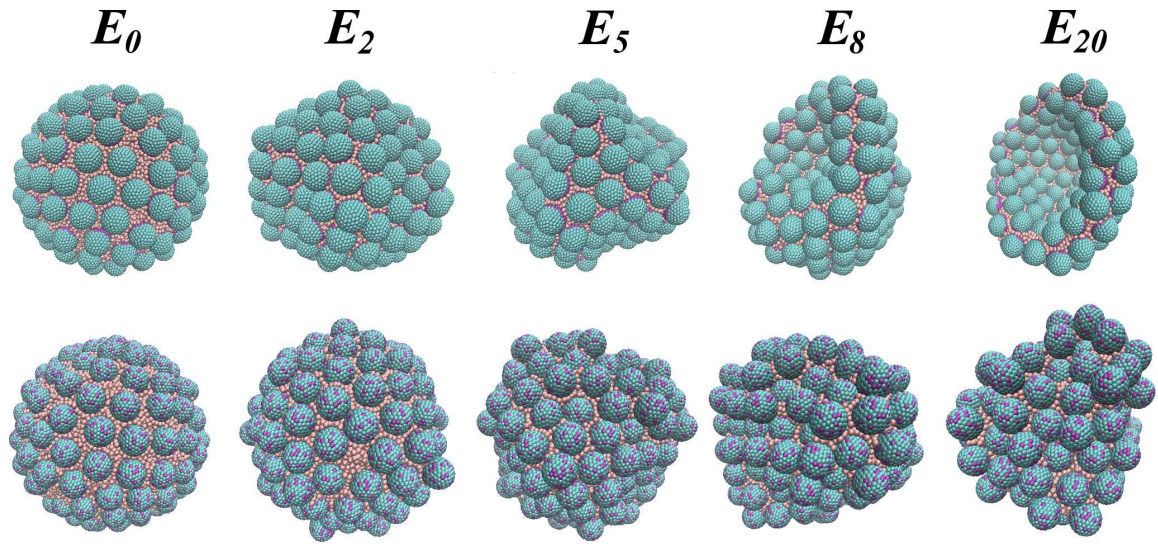}
 \caption{
 Sequence of simulation snapshots representing buckling processes of water in oil 
 droplets armored with 160 spherical Janus (top) and homogeneous (bottom) nanoparticles after 
 successive removals of water. The number of water beads removed increases from left to right 
 with $E_i$ refering to the $i^{th}$ removal. Cyan and purple spheres represent polar 
 and apolar beads, respectively. Pink spheres represent water beads. The oil molecules surrounding the 
 system are not shown for clarity.
}
\label{fig1}
\end{figure*}

We employ here Dissipative Particle Dynamics (DPD)~\cite{Groot1997} as a mesoscopic simulation method. We study 
the shape and buckling transitions of model water droplets coated with spherical nanoparticles and immersed in an 
organic solvent. The procedure and the parametrisation details are fully described in prior 
work \cite{LuuLangmuir2013, LuuJPCB2013, SicardFD2016} and in the Supporting Information (SI). 
The particles are of two different types: Janus and homogeneous. They are chosen so that the 
initial three-phase contact angles ($\approx 90^\circ$) result in maximum adsorption energy.
The volume of the droplets is controllably reduced, pumping randomly a constant proportion of water molecules 
out of the droplet (more details in the SI). At every stage we remove $10$ percent of the water from the droplet. Throughout this letter, $E_i$ refers to the $i^{th}$ removal of water, with $E_0$ corresponding to 
the initial configuration and $E_{20}$ to the final configuration. We seek to determine whether the NPs 
at the droplet interface buckle,  causing the droplets to deviate from the spherical shape. 
We show that Janus particles affect strongly the shape of the droplet via the formation of 
a crater-like depression. This evolution is \textit{actively} controlled by a close-packed particle 
monolayer at the curved interface. On the other hand, homogeneous particles follow \textit{passively} 
the volume reduction of the droplet. The shape of the droplet remains approximately spherical 
with a nanoparticle monolayer/bilayer transition, with some NPs desorbing in water.
We discriminate the two mechanisms with the evolution of their respective nanoparticle three-phase 
contact angle distributions. While for Janus particles the distribution remains unimodal, albeit skewed 
when the droplet is significantly shrinked, 
for homogeneous particles, the evolution of the contact angle distribution becomes bimodal with some particles 
becoming more/less immersed in the aqueous phase.

\begin{figure*}
\includegraphics[width=0.85\textwidth, angle=-0]{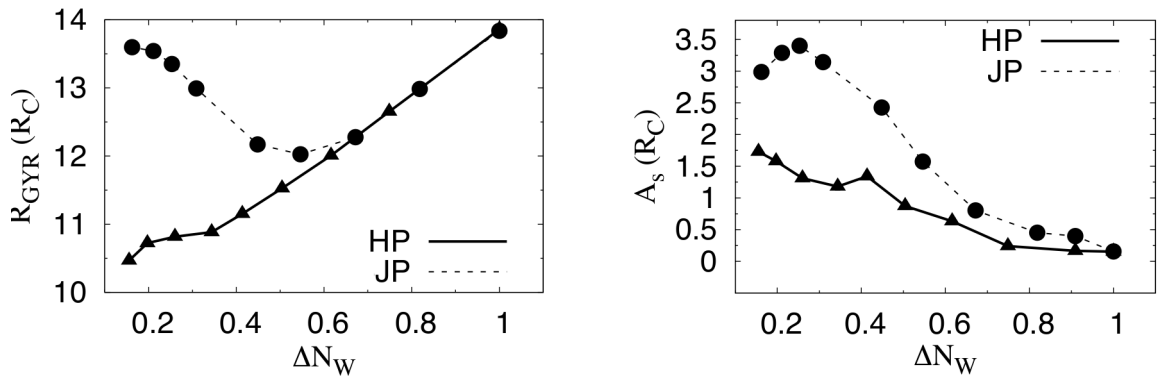}
 \caption{Temporal evolution of the radius of gyration, $R_{\textrm{GYR}}$ (left panel) 
 and the asphericity $A_{s}$ (right panel), for armored droplets stabilized  
 with Janus (circles and dashed line) and homogeneous NPs (triangles and plain line) 
 as a function of the dimensionless parameter $\Delta N_W \equiv N_W/N_W^0$. 
 $N_W$ represents the number of water beads that remain in the droplet after each removal, and 
 $N_W^0$ is the initial number of water beads. The statistical errors 
 are estimated as one standard deviation from the average obtained for equilibrated trajectories, 
 and they are always smaller than the symbols.
 For comparison with snapshots in Fig.~\ref{fig1}, $\Delta N_W(E_2) \approx 0.82$, $\Delta N_W(E_5) \approx 0.62$, 
 $\Delta N_W(E_8) \approx 0.45$, and $\Delta N_W(E_{20}) \approx 0.16$.}
\label{fig2}
\end{figure*}

We consider a system initially made by a spherical water droplet immersed in oil, 
and stabilized by a sufficiently dense layer of NPs \cite{SicardFD2016}. The initial shape 
of the droplet is spherical. The only difference between the two systems is the NP chemistry, 
\textit{i.e.} the distribution and proportion of polar and apolar beads around the spherical 
particles and their efficiency in interacting with the two fluids at the interface.
Janus and homogeneous NPs are designed to present comparable three-phase contact angles, 
$\theta_c = (91.6 \pm 2.0)^\circ$ and $\theta_c = (88.7 \pm 3.5)^\circ$, respectively (\textit{cf.} SI for details). 
We consider throughout this study the same NP density on the droplets.  
We calculate the radius of gyration, $R_{\textrm{GYR}}$, and the asphericity, $A_s$, for the droplet covered by either
Janus or  homogeneous NPs (\textit{cf.} SI for details). For the initial configurations, we obtain 
$R_{\textrm{GYR}} = 13.837 \pm 0.003$ and $R_{\textrm{GYR}} = 13.860 \pm 0.003$, 
and $A_s = 0.156 \pm 0.05$ and $A_s = 0.153 \pm 0.05$, 
respectively, expressed in $R_C$ units (\textit{cf.} SI for details).

In Fig.\ref{fig1} we show representative snapshots obtained during the simulations for systems containing 
Janus NPs (top panels) and homogeneous NPs (bottom panels). 
Visual inspection of the simulation snapshots highlights some fundamental 
differences between the two buckling processes. 
We start with spherical initial droplets ($E_0$).
When the water droplet is coated with Janus particles (top), the system starts developing dimples as 
moderate amount of water is removed ($E_2$). The morphology then becomes more crumpled with increasing numbers 
of dimples ($E_5$). 
For stronger removal, the droplet geometry evolves to a large and smooth curved shape, yielding 
a crater-like depression to minimize the interfacial energy of the system ($E_8$ and $E_{20}$). 
During this evolution, Janus NPs remain strongly adsorbed at the interface, forming 
a close-packed monolayer between the two fluids.

The buckling process is fundamentally different when the water droplet is stabilized with 
homogeneous NPs (bottom). When the volume of the droplet is reduced, the shape of the system 
evolves smoothly and does not present any sharp transition to morphologies showing dimples and cups, 
nor crater-like depressions. Instead, the NPs reorganize progressively into a bilayer, 
presumably to minimize the system energy. Unlike Janus NPs, homogeneous NPs either protrude exceedingly towards 
the decane solvent, or recede into the water droplet with some particles even desorbing into the water phase (from $E_2$ to $E_{20}$). 
For reference, we recall that the change in energy accompanying desorption of a spherical particle from 
the oil-water interface to either bulk phase is approximated by $\Delta E = \pi r^2 \gamma_{ow} (1 \pm \cos \theta)^2$, 
in which $r$ is the particle radius, $\gamma_{ow}$ is the bare oil-water interfacial tension, and the plus (minus) sign 
refers to desorption into oil (water)~\cite{BinksSM2016}. 
Even if this expression assumes 
the oil-water interface remains planar up to the contact line with the particle, it can give a rough approximation 
of the energy at play. Considering the system parameters given in the SI, 
we obtain $\Delta E \approx 85~k_BT$ in our systems when one NP desorbs.

These two different behaviours are quantitatively investigated in Fig.~\ref{fig2}, where we show 
the temporal evolution of the radius of gyration (left panel), $R_{\textrm{GYR}}$, and the asphericity 
(right panel), $A_s$, of the two droplets 
as a function of the dimensionless parameter $\Delta N_W \equiv N_W/N_W^0$, 
with $N_W$ the number of water beads remaining in the droplet, and 
 $N_W^0$ the initial number of water beads in the droplet.
When $N_W > 0.6$, the radius of gyration of two systems follow the same evolution, 
regardless the chemistry of the NPs (Janus or homogeneous). 
For one droplet coated with Janus NPs, $R_{GYR}$ then departs from its linear trend 
when $N_W < 0.6$. This departure corresponds to the evolution from $E_5$ to $E_8$ in Fig.~\ref{fig1}, 
\textit{i.e.} the transition from a droplet interface made of dimples and cups to the formation of 
the crater-like depression. 
During this transition, the size of one dimple increases when the system relaxes after evaporation. 
This local evolution  yields a larger depression, which causes the progressive coalescence  
of the small dimples.
This transition is consistent with the \textit{surface model} numerical analysis from Ref.~\cite{QuillietEPJE2012}, 
which study the shape evolution of a spherical elastic surface when the volume it encloses is decreased. 
This model, which has long been considered as valid to describe the deformation of thin shells
~\cite{LandauBook1986,AudolyBook2010}, showed that a thin shell with a single dimple has 
lower energy than a shell containing  multiple dimples. This occurs because elastic energy mainly concentrates 
in dimple edges as bending energy. Dimples coalescence lowers the total elastic energy.
Below $\Delta N_W \approx 0.6$, $R_{GYR}$ increases as the droplet can be described as half-sphered.
Let us note that this evolution is coherent with the temporal evolution of the radial distribution function 
of the NPs, $g(r)$, with $r$ the distance between the centers of the NPs, given in the SI.
In contrast, the droplet coated with homogeneous NPs shrinks isotropically when the volume reduces 
even below $\Delta N_W \approx 0.6$. 
This evolution yields continuous decrease of $R_{GYR}$ and 
a relatively low $A_s$ in Fig.~\ref{fig2}. Eventually, the NP concentration becomes too high 
and some NPs move into the droplet. 
When $N_W < 0.25$, the number of water beads that remain in the droplet is not sufficient to define unambiguously the droplet volume. This limitation impacts the system shape and 
the evolution of $R_{GYR}$ and $A_s$ for Janus and homogeneous NPs.
\begin{figure*}
\includegraphics[width=0.9\textwidth, angle=-0]{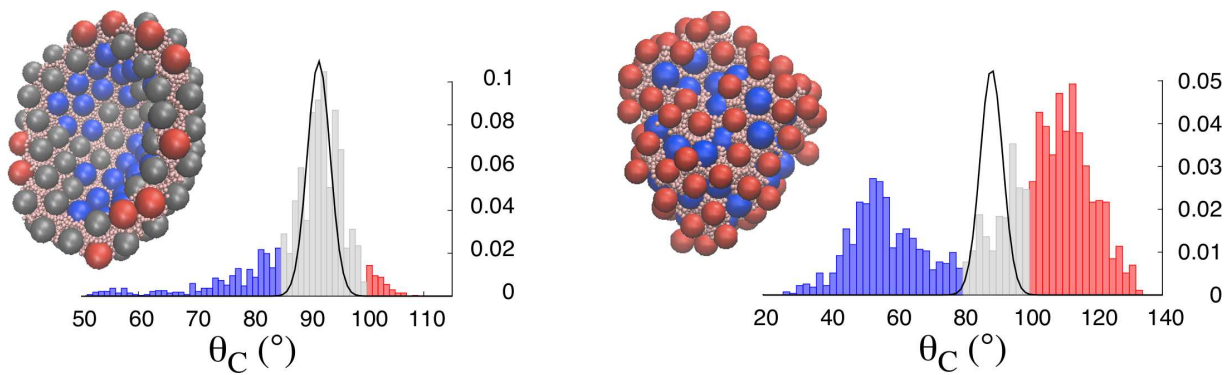}
 \caption{Three-phase contact angle distribution of Janus (left panel) 
and homogeneous (right panel) NPs at the initial stage $E_0$ (continuous black lines) where the shape 
of the droplet is spherical, and at the final stage $E_{20}$ (histograms). The initial distributions (stage $E_0$) 
are fitted with  Gaussian distributions for both systems.  
The droplet configurations at the final stage $E_{20}$ are also shown. 
The blue, gray, and red spheres represent the NPs with three-phase contact angles in the three respective 
regions highlighted in the histogram distributions.}
\label{fig3}
\end{figure*}

We also quantify the NP layer properties to see if the NPs actively influence or passively follow 
the evolution of the droplet geometry.
In Fig.~\ref{fig3}, we compare the three phase contact angle distribution of Janus (left panel) 
and homogeneous (right panel) NPs at the initial stage $E_0$, where the shape of the droplet is spherical, 
and the final stage, $E_{20}$. The initial distributions, fitted with  continuous lines, 
can be described with  Gaussian distributions for both NPs.
The values of the respective means, $\mu^J$ and $\mu^H$, and 
variances, $\sigma^J$ and $\sigma^H$, differ due to the NPs chemistry. We obtain 
$\mu^J = 91.6^\circ$ and $\mu^H = 88.6^\circ$, and $\sigma^J=2.0^\circ$ and $\sigma^J=3.4^\circ$ 
for Janus and homogeneous NPs, respectively.
When the droplet coated with Janus NPs shrinks, the contact angle distribution evolves to a skewed one, 
but it remains unimodal, with a single peak centered at the same value as the one measured for the  
initial configuration. The emergence of the skewness of the distribution is linked to the decrease 
of the NP-NP distance when the droplet volume is reduced. It is due to the major role played by steric effects.
As discussed earlier, to minimize its interfacial energy, the system must deform its shape, eventually 
forming a crater-like depression. We conclude this transition is achieved through the \textit{active role} 
played by the Janus NPs.
In the final structure, some NPs are forced to deviate from their original contact angle, increasing the skewness 
of the distribution on both sides of the peak. 

The evolution of the system is different when homogeneous NPs are present. As the droplet volume is reduced, 
the contact angle distribution firstly evolves as a monolayer interface with a single peak  
(cf. SI). As the droplet shrinks further, and the distance between the NPs decreases, 
the distribution becomes bimodal, with two distinct peaks emerging on both sides of the initial equilibrium 
contact angle. 
This feature is characteristic of a particle bilayer. Indeed, homogeneous NPs are more weakly attached 
to the interface than Janus NPs. In the case of buckling mechanism studied here, the homogenous NPs mainly follow the 
volume reduction, sharing the interfacial area, either receding into the water droplet 
or protruding towards the organic solvent. 
Unlike Janus NPs, homogeneous NPs do not drive 
the evolution of the droplet shape, which does not differ too much from the spherical geometry. 
The behaviour just described is characteristic of the \textit{passive role} played by the homogeneous NPs, 
which mainly follow the volume reduction, only modulating the droplet shape due to the steric constraints. 

The curved shape obtained when the droplet is coated with Janus NPs can also be characterized by 
the wettability associated with the local arrangement of the NPs at the interface.
The particles with a contact angle $\theta_c < 85^\circ$, \textit{i.e.} the blue ones in Fig.~\ref{fig3} 
(left panel), can be found in the crater-like depression. These particles have receded  
into the water droplet due to the concave local geometry of the interface. The particles with $\theta_c > 100^\circ$,
 \textit{i.e.} the red ones in Fig.~\ref{fig3} (left panel), can be found at the 
transition between the concave and convex areas of the interface, where they are likely to protrude towards 
the solvent.
The shape deformation of the droplet is achieved through the \textit{active role} played by the 
Janus NPs. Their specific chemistry causes them to create an interface with excess wettability
($\theta_c < 85^{\circ})$ in a \textit{pocket} delimited by the crater-like depression, and surrounded by a cup with low wettability ($\theta_c >100^{\circ}$).

\begin{figure*}
\includegraphics[width=0.7\textwidth, angle=-0]{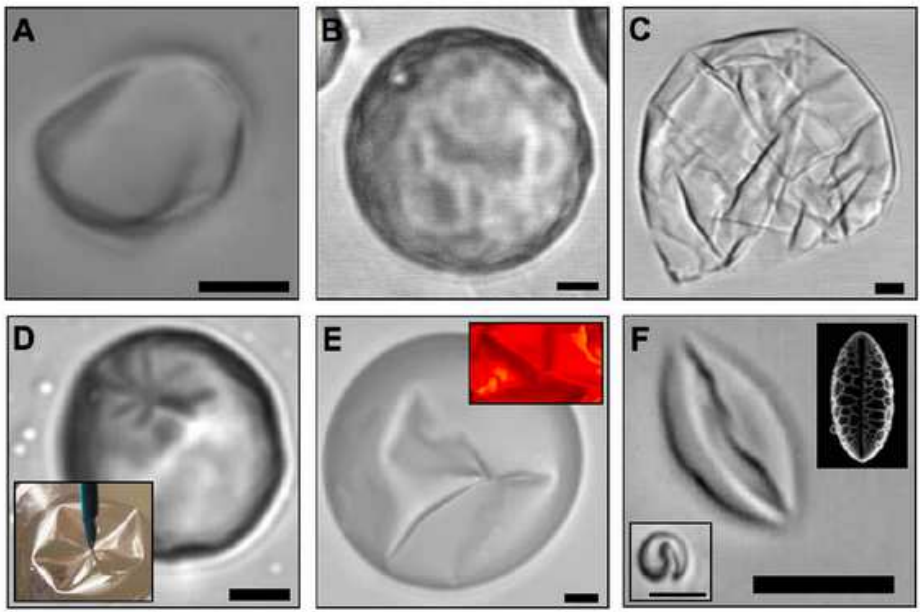}
 \caption{Optical micrographs of buckled droplets obtained experimentally  
 by Datta at al.~\cite{DattaLangmuir2010}. Panels (A-C) show characteristic shapes at increasing levels of evaporation, 
 and panels (D-F) show typical buckled structures (\textit{cf.} Ref.~\cite{DattaLangmuir2010} for experimental details).
  All scale bars are $5~\mu m$. Datta et al. used hydrophilic silica NPs coated with a diffuse 
  layer of alkane, rendering them partially hydrophobic and partially hydrophilic. The resulting three-phase 
  contact angle in Ref.~\cite{DattaLangmuir2010} was $\approx 90^\circ$.}
\label{fig4}
\end{figure*}

Our results are consistent with experiments reporting buckling and crumpling of nanoparticle-coated droplets~\cite{XuLangmuir2005,TsapisPRL2005, DattaLangmuir2010}.
In particular, we observe a close analogy to the experimental work of Datta et al.~\cite{DattaLangmuir2010}, 
who studied water-in-oil droplets of varying sizes. In these experiments the dispersed phase is slightly soluble 
in the continuous phase. The volume reduction was controlled with the addition of a fixed amount of unsatured continous phase.
As shown in Fig.~\ref{fig4}, Datta et al. observed droplet shapes including dimples, cups, and folded configurations, 
in agreement with our simulations (cf. experimental details in the caption in Fig.~\ref{fig4}).
Unlike our mesoscopic analysis, Datta et al.~\cite{DattaLangmuir2010} do not have access 
to the particle three-phase contact angle distribution. This information provides a deeper understanding 
of the organisation of the NPs at the interface, and allows us to decipher the \textit{active} 
or \textit{passive} role of the NPs. 

As explained in the SI, the layering properties of the particles 
depend strongly on the numerical protocol. For example, decreasing the relaxation 
time between successive water removals can induce NP release 
from the interface, which is in agreement with experiments~\cite{GarbinLangmuir2011}. 
The results presented here seem to be due to the chemistry of the nanoparticles simulated 
(\textit{i.e.} Janus \textit{vs.} homogeneous). It is however possible that homogeneous NPs 
with large adsorption energy become \textit{active} and yield buckled armored droplets similar 
to those observed when Janus NPs are simulated here.

The new physical insights discussed in this letter could be useful for a variety of applications. 
For example, controlling the positions of the solid particles with respect to the interface could 
help in heterogeneous catalysis~\cite{CrossleyScience2010}.
In biomimetic design, where the identification and evaluation of surface binding-pockets is crucial,
the ability of controlling pockets such as those created by the crater-like depression in the 
presence of Janus NPs, could play a central role 
in designing structures with a defined geometry~\cite{PatelJPCB2014}. 
The analogy between Fig.~\ref{fig1} and the shape of protein active site might play 
an important role for ligand docking~\cite{WeiselCCJ2007,MarcosScience2017}. 
Finally, buckled armored droplets might also be of relevance as potential drug delivery systems~\cite{FrelichowskaIJP2009}. 
Over the last decade, nanoscale droplets have been used  
for instant real-time ultrasound imaging of specific organs~\cite{LeeAM2015}. Superparamagnetic 
solid NPs provide a means of manipulating the droplets using an external magnetic field~\cite{LeeAM2015}. 
One of the main limitations in such applications is droplet coalescence, which can happen before droplets 
reach the target. 
The specific shapes obtained with buckled armored droplets might prevent  coalescence. Indeed, the NP arrangements on the droplets show increased  packing, which reduces significantly the NPs mobility. The particle layers would
then provide enough mechanical resistance to guarantee the droplet stability.

\section*{acknowledgements}

The authors acknowledge V. Garbin, and L. Botto for useful discussions. Via our membership of 
the UK's HEC Materials Chemistry Consortium, which is funded by EPSRC (EP/L000202), 
this work used the ARCHER UK National Supercomputing Service (http://www.archer.ac.uk). 
F.S. is supported by the UK Engineering and Physical Sciences Research Council (EPSRC), 
under grant number 527889.

\bibliography{prl} 
\bibliographystyle{apsrev4-1} 
%
\pagebreak
\widetext
\begin{center}
\textbf{\large Buckling in Armored Droplets \\
			   Supporting Information}
\end{center}
\setcounter{equation}{0}
\setcounter{figure}{0}
\setcounter{table}{0}
\setcounter{page}{1}
\makeatletter
\renewcommand{\theequation}{S\arabic{equation}}
\renewcommand{\thefigure}{S\arabic{figure}}
%
\section{MD simulation method}
The Dissipative Particle Dynamics (DPD) simulation method~\cite{Groot1997} was implemented within 
the simulation package LAMMPS~\cite{Plimpton1995}. The procedure and the parametrisation details 
are fully described in prior work~\cite{LuuLangmuir2013, LuuJPCB2013}. The system simulated here 
is composed of water, oil (decane), and nanoparticles (NPs). One "water bead" (w) represents 5 water molecules 
and a reduced density of one DPD bead is set to $\rho=3$. One decane molecule is modeled as two "oil beads" (o) 
connected by one harmonic spring of length $0.72$ $R_c$ and spring constant $350$ $k_BT/R_c$~\cite{Groot2001}, 
where $R_c$ is the DPD cutoff distance. The initial size of the simulation box 
is $L_x \times L_y \times L_z \equiv 72 \times 72 \times 78$ $R_c^3$, where $L_i$ is the box length 
along the $i^{th}$ direction. Periodic boundary conditions are applied in all three directions.
The NPs are modelled as hollow rigid spheres and contain polar (p) and nonpolar (ap) DPD beads 
on their surface. One DPD bead was placed at the NP center for convenience, as described 
elsewhere~\cite{LuuLangmuir2013, LuuJPCB2013}. Hollow models have been used in the literature to simulate NPs, 
and hollow NPs can also be synthesized experimentally~\cite{Calvaresi2009}. 
We considered spherical NP of the same volume, $4/3 \pi a_0^3$, where $a_0$ is the radius of the sphere. 
We imposed $a_0 = 2R_c \approx 1.5$ nm.
All types of beads in our simulations have reduced mass of $1$. We maintain the surface bead density on the NPs
sufficiently high to prevent other DPD beads (either decane or water) from penetrating the NPs 
(which would be unphysical), as it has already been explained elsewhere~\cite{LuuJPCB2013}. 
To differenciate every NPs, we report the nonpolar fraction of the NP surface beads and the NP type. 
For example, $75HP$ ($55JP$) indicates that $75\%$ ($55\%$) of the beads on the NP surface are nonpolar, 
and that we consider an homogeneous (Janus) NP.\\

The interaction parameters shown in Table \ref{Tab-interaction} are used here. These parameters were 
adjusted to reproduce selected atomistic simulation results, as explained in prior work~\cite{LuuLangmuir2013}. 
By tuning the interaction parameters between polar or nonpolar NP beads and the water and decane beads 
present in our system, it is possible to quantify the effect of surface chemistry on the structure 
and dynamics of NPs at water-oil interfaces. Specifically, the interaction parameters between NP polar 
and nonpolar beads were adjusted to ensure that NPs are able to assemble and disassemble without yielding 
permanent dimers at the water/oil interface~\cite{LuuLangmuir2013}.
All simulations were carried out in the NVE ensemble~\cite{LuuLangmuir2013}. The scaled temperature was $1$, 
equivalent to $298.73$ K. The DPD time scale can be gauged by matching the self-diffusion of water. 
As demonstrated by Groot and Rabone~\cite{Groot2001}, the time constant of the simulation can be calculated 
as $\tau = \frac{N_m D_{\textrm{sim}} R_c^2}{D_{\textrm{water}}}$, where $\tau$ is the DPD time constant, $D_{\textrm{sim}}$ is the simulated water self-diffusion coefficient, and $D_{\textrm{water}}$ is the experimental 
water self-diffusion coefficient. When $a_{w-w} = 131.5$ $k_B T/R_c$ (cf. Tab.~\ref{Tab-interaction}), 
we obtained $D_{\textrm{sim}} = 0.0063$ $R_c^2/\tau$. For $D_{\textrm{water}} = 2.43 \times 10^{-3}$ $cm^2/s$~\cite{Partington1952}, we finally obtain $\tau = 7.6$ ps.\\
\begin{table}[h]
\begin{center}
\begin{tabular*}{0.45\textwidth}{@{\extracolsep{\fill}}ccccc}
  \hline
  {} & $w$ & $o$ & $ap$ & $p$ \\
  \hline
  $w$ & $131.5$ & $198.5$ & $178.5$ & $110$  \\
  $o$ & {} & $131.5$ & $161.5$ & $218.5$ \\
  $ap$ & {} & {} & $450$ & $670$ \\
  $p$ & {} & {} & {} & $450$ \\
  \hline
\end{tabular*}
\caption{DPD interaction parameters expressed in $k_BT/R_c$ units. Symbols $w$, $o$, $ap$, and $p$ stand for 
water beads, oil beads, NP nonpolar beads, and NP polar beads, respectively.}
\label{Tab-interaction}
\end{center}
\end{table}

\section{System characterisation: Droplets and Nanoparticles}
In our simulations, the initial size of the droplet was fixed. At the beginning 
of each simulation, the solvent (oil) beads were distributed within the simulation box forming 
a cubic lattice. One water droplet of radius $\approx 20$ $R_c$ was generated by replacing the oil beads 
with water beads within the volume of the spherical surface. 
A number of spherical NPs were placed randomly at the water-decane interface with their polar (nonpolar) 
part in the water (oil) phase to reach the desired water-decane interfacial area per NP. 
Following previous work~\cite{LuuJPCB2014,SicardFD2016}, the NPs considered in this study are spherical 
and of two different types: Janus and homogeneous. The emulsion systems are stabilized by a sufficiently 
dense layer of NPs~\cite{SicardFD2016}. We considered water droplets coated with $160$ 
spherical Janus and homogeneous nanoparticles of type $55JP$ and $75HP$, respectively. Considering the NP surface coverage, $\phi$, defined in Ref.~\cite{LuuLangmuir2013}, 
we obtain $\phi \approx 0.9$. Considering the results obtained in Ref.~\cite{LuuLangmuir2013} 
for a flat interface, this yields an interfacial tension $\gamma_{ow} \approx 6.8$ $k_BT/R_c^2$ 
for both Janus and homogeneous NPs.
The initial configuration obtained was simulated for $10^6$ timesteps in order to relax the density of 
the system and the contact angle of the nanoparticles on the droplet. The system pressure 
and the three-phase contact angles did not change notably after 5000 simulation steps.
We let then run the system for an additional $2\times 10^6$ timesteps to generate two new intial configurations 
after $2\times 10^6$ and $3\times 10^6$ timesteps, respectively. We then repeated the buckling simulation 
with these different initial configurations to test the reproducibility of the simulation.\\

The surface area of the droplets is \textit{slowly} diminished, pumping randomly a constant proportion, 
\textit{i.e.} $10$ percent, of water molecules out of the droplet and letting the system pressure 
and the three-phase contact angles equilibrate at constant density. By \textit{slowly}, 
we mean we do not create any hollow volume in the droplet that would strongly drive the system 
out-of-equilibrium. Doing so, the three-phase contact angle distribution of the NPs 
evolves sufficiently smoothly when the droplet buckles and becomes nonspherical, thereby preventing 
particles to be \textit{artifactually} realeased. This numerical protocol can be similarly compared with 
an emulsion system where the dispersed phase is slightly soluble in the continuous phase~\cite{DattaLangmuir2010}. 
By adding a fixed amount of unsatured continuous phase, the volume of the droplets can then be controllably reduced.\\

\textbf{Three-phase contact angle}. To estimate the three phase contact angle on the droplets we calculate 
the fraction of the spherical NP surface area that is wetted by water~\cite{FanSM2012},
\begin{equation}
\theta_C = 180 - \arccos\Big(1-\frac{2 A_w}{4\pi R^2}\Big) \, ,
\end{equation}
where $A_w$ is the area of the NP surface that is wetted by water and $R$ is the radius of the NP. The ratio $A_w/4\pi R^2$ 
is obtained by dividing the number of NP surface beads (ap or p), which are wetted by water, by the total number of beads 
on the NP surface ($192$ for spherical NP). One surface bead is wet by water if a water bead is the solvent bead nearest to it. 
One standard deviation from the average is used to estimate the statistical uncertainty. \\

\textbf{Radius of gyration}. The description of the geometrical properties of complex systems 
by generalized parameters such as the radius of gyration or principal components of the gyration 
tensor has a long history in macromolecular chemistry and biophysics~\cite{Vymetal2011,Solc1971,SicardFD2016}.
Indeed, such descriptors allow an evaluation of the overall shape of a system and reveal its symmetry. 
Considering, e.g., the following definition for the gyration tensor, 
\begin{equation}
\mathcal{T}_{GYR} = \frac{1}{N}
\begin{bmatrix}
\sum x_i^2 & \sum x_i y_i & \sum x_i z_i \\
\sum x_i y_i & \sum y_i^2 & \sum y_i z_i \\
\sum x_i z_i & \sum y_i z_i & \sum z_i^2
\end{bmatrix} \, ,
\end{equation}
where the summation is performed over $N$ atoms and the coordinates $x$, $y$, and $z$ are related to the 
geometrical center of the atoms, one can define a reference frame where $\mathcal{T}_{GYR}$ can be diagonalized:
\begin{equation}
\mathcal{T}_{GYR}^{diag} =
\begin{bmatrix}
S_1^2 & 0 & 0\\
0 & S_2^2 & 0 \\
0 & 0 & S_3^2
\end{bmatrix} \, .
\end{equation}
In this format we obey the convention of indexing the eigenvalues according to their magnitude. 
We thus define the radius of gyration $R_{GYR}^2 \equiv S_1^2 + S_2^2 + S_3^2$, and the asphericity 
$A_s \equiv S_1 - \frac{1}{2}(S_2+S_3)$, which measures the deviation from the spherical symmetry. 
To determine the properties of a droplet, we calculate $R_{GYR}$ and $A_s$ using the centers of the water beads.\\

In this letter, Janus and homogeneous NPs present similar three-phase contact angle $\theta_c = (91.6 \pm 2.0)^\circ$ 
and $\theta_c = (88.7 \pm 3.5)^\circ$, respectively. The radius of gyration, $R_{\textrm{GYR}}$, 
and the asphericity, $A_s$, for the Janus and homogeneous initial configurations are 
$R_{\textrm{GYR}} = 13.837 \pm 0.003$ and $R_{\textrm{GYR}} = 13.860 \pm 0.003$, 
and $A_s = 0.156 \pm 0.05$ and $A_s = 0.153 \pm 0.05$, respectively, expressed in $R_C$ units.

\section{Evolution of the Nanoparticle radial distribution function}
\begin{figure*}
\includegraphics[width=0.55\textwidth, angle=-0]{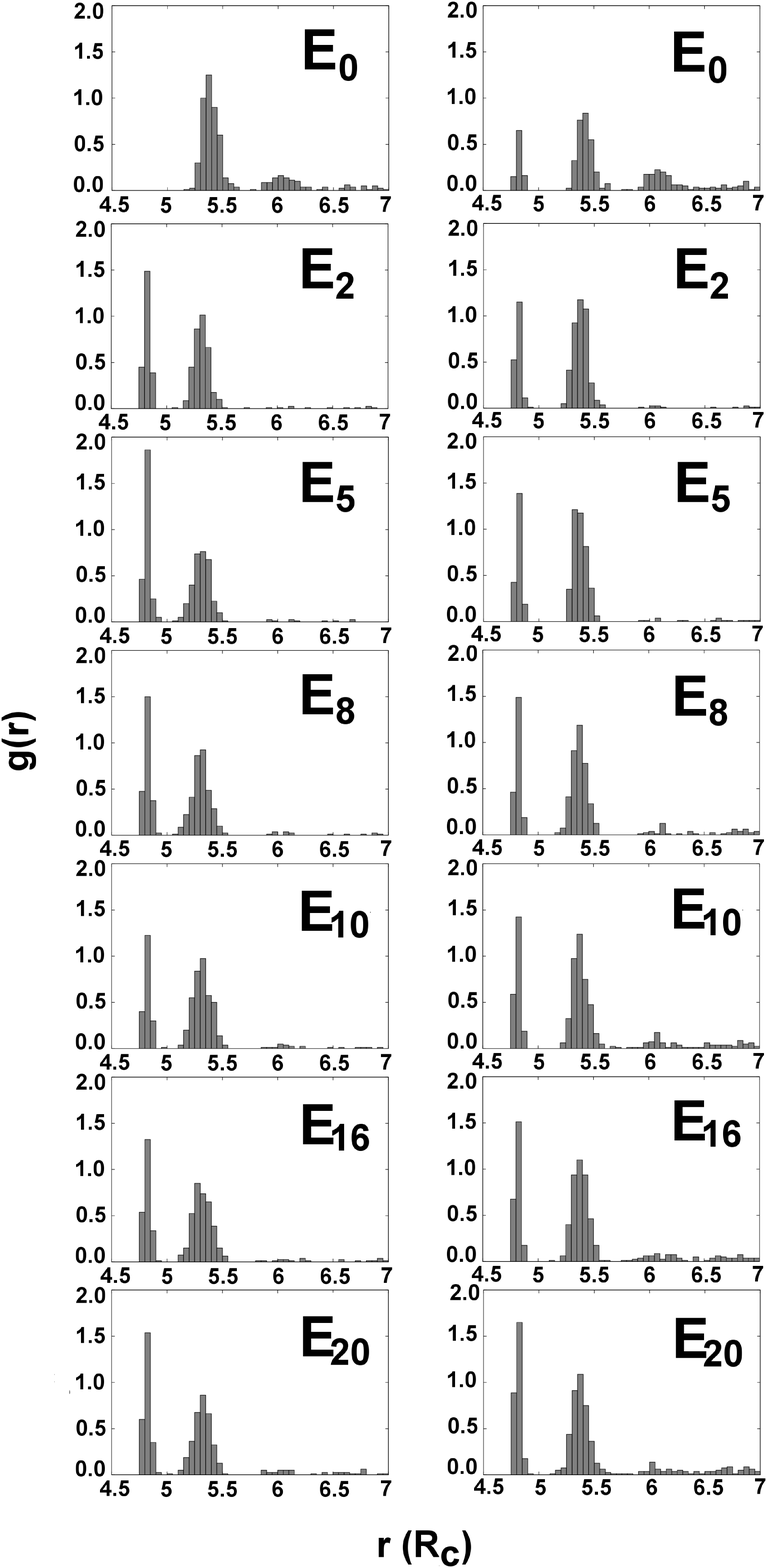}
 \caption{Evolution of the radial distribution function of the NPs, $g(r)$. $r$ is  
 the distance between the centers of the NPs. The water droplet is coated with Janus (left) 
 and homogeneous (right) particles. For comparison with snapshots in Fig.~1 in the main text, 
 $E_i$ refers to the $i^{th}$ removal of water beads. We follow the evolution of the first and second neighbouring shells. }
\label{figS1}
\end{figure*}

The transition from dimples and cups to crater-like depression observed when Janus nanoparticles 
cover the droplet can be reflected in the temporal evolution of the radial distribution function of the NPs, $g(r)$, 
with $r$ the distance between the centers of the NPs. 
We first consider the initial spherical configuration, and extract the list 
of nearest neighbours of each NP within a shell of radius $r < 12R_c$. This treshold value defines 
the first and second neighbouring shells~\cite{FanPRE2012,SicardFD2016}.
As the system is densely packed at the interface, we follow the temporal evolution of $g(r)$ 
considering this subset of particles, which corresponds to the nearest neighbours shell.
When the volume of the droplet reduces, we first see in Fig.~\ref{figS1} (left panel) the emergence 
of a new peak around $r \approx 4.75 R_C < r_{\textrm{DPD}}$, where $r_{\textrm{DPD}} = 5 R_C$ represents 
the distance between the centers of NPs above which the NPs do not interact with each other through 
the DPD non-bonded force.
As the droplet volume further reduces, the heigh of the new peak increases. 
This evolution continues until $\Delta N_W \approx 0.6$ (cf. Fig.~2 in the main text). 
Below that value, the heigh of the first peak decreases, together with the increase of the heigh 
of the second peak. This corresponds to the transition from dimples and cups to crater-like depression. 
The evolution continues until $\Delta N_W \approx 0.25$ where the number 
of water beads which remain in the droplet is not sufficient to define unambiguously the droplet volume.\\

As the $g(r)$ evolution between the homogeneous particles is concerned, 
we see in Fig.~\ref{figS1} (right panel), a constant increase of the heigh of the first peak, centered 
 at $r \approx 4.75 R_C$. This first peak located below $r_{\textrm{DPD}}$ 
is already present in the initial spherical configuration, and is linked to the difference between 
the NP chemistry.
Indeed, Janus NP are made of two hemispherical faces, one hydrophobic in contact with 
the hydrocarbon beads, and one hydrophilic in contact with the water beads. 
The NP hydrophobic faces which protrude in the solvent interact strongly and repeal each other 
to minimize the interaction potential energy. 
Unlike Janus NPs, both hemispherical faces of homogeneous NPs are made with hydrophobic 
and hydrophilic beads. This allows the NPs to interact smoothly with each other and to fluctuate more. 
Hence they can come closer than $r_{\textrm{DPD}}$, keeping spherical the droplet interface.

\section{Evolution of the three-phase contact angle distribution}
\begin{figure*}
\includegraphics[width=0.90\textwidth, angle=-0]{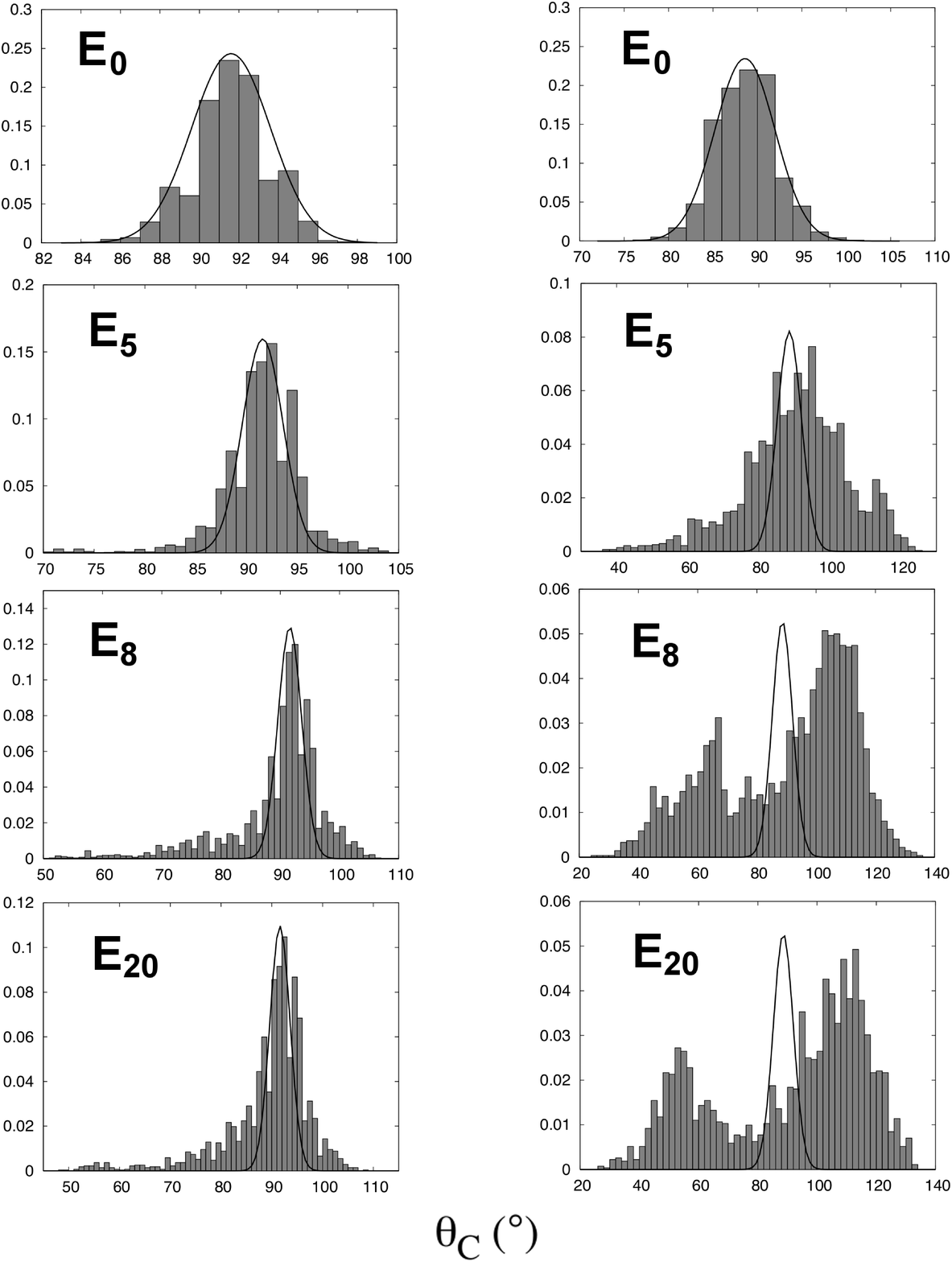}
 \caption{Evolution of the three-phase contact angle distribution, $\theta_C$, of Janus (left panels) 
 and homogeneous (right panels) NPs. For comparison with snapshots in Fig.~1 in the main text, 
 $E_i$ refers to the $i^{th}$ removal of water beads. The initial distributions (stage $E_0$) are fitted 
 with Gaussian distributions, which are plotted at every stage, 
 for comparison. For clarity, we rescale the height of the Gaussian distributions at every stage.}
\label{figS2}
\end{figure*}

In Fig.~\ref{figS2}, we show the evolution of the three phase contact angle distribution of Janus 
(left panel) and homogeneous (right panel) NPs from the initial stage $E_0$ (cf. Fig.~1 in the main text), 
where the shape of the droplet is spherical, to the final stage $E_{20}$. 
The initial distributions, fitted with continuous lines, can be described with  Gaussian distributions 
for both the Janus and homogenous emulsion systems. The values of the respective means, $\mu^J$ and $\mu^H$, 
and variances, $\sigma^J$ and $\sigma^H$, differ according to the chemistry of the stabilizers. We obtain 
$\mu^J = 91.6^\circ$ and $\mu^H = 88.6^\circ$, and $\sigma^J=2.0^\circ$ and $\sigma^J=3.4^\circ$ 
for Janus and homogeneous NPs, respectively.\\

When the droplet is coated with Janus NPs, the system evolves to a skewed distribution as the droplet shrinks, 
but remains unimodal with a single peak at the same value as the one measured for the spherical 
initial configuration. The emergence of the skewness of the distribution is linked to the decrease 
of the NP-NP distance when the droplet volume is reduced due to the major role played by the steric effect.
The evolution is different when homogeneous NPs cover the droplet. As the volume is reduced, 
the contact angle distribution firstly evolves as a monolayer interface with a single peak 
(from $E_{0}$ to $E_{5}$, Fig.~1 in the main text). 
As the distance between the NPs decreases further, the distribution becomes bimodal as two distinct peaks emerge 
on both sides of the original equilibrium contact angle. This fundamental difference is characteristic of 
a particle bilayer. 

\newpage

\section{How numerical algorithms affect the droplet evolution}
In our mesoscopic analysis, we controllably reduced the volume of the droplet, removing a 
constant proportion, \textit{i.e.} 10 percent, of water beads from the droplet. As a consequence, the three-phase 
contact angle distribution of the NPs evolves smoothly when the droplet buckles, thereby preventing particles to be artifactually released.
However, the particles behaviour strongly depends on the numerical protocol.\\

\begin{figure}
\includegraphics[width=0.7\textwidth, angle=-0]{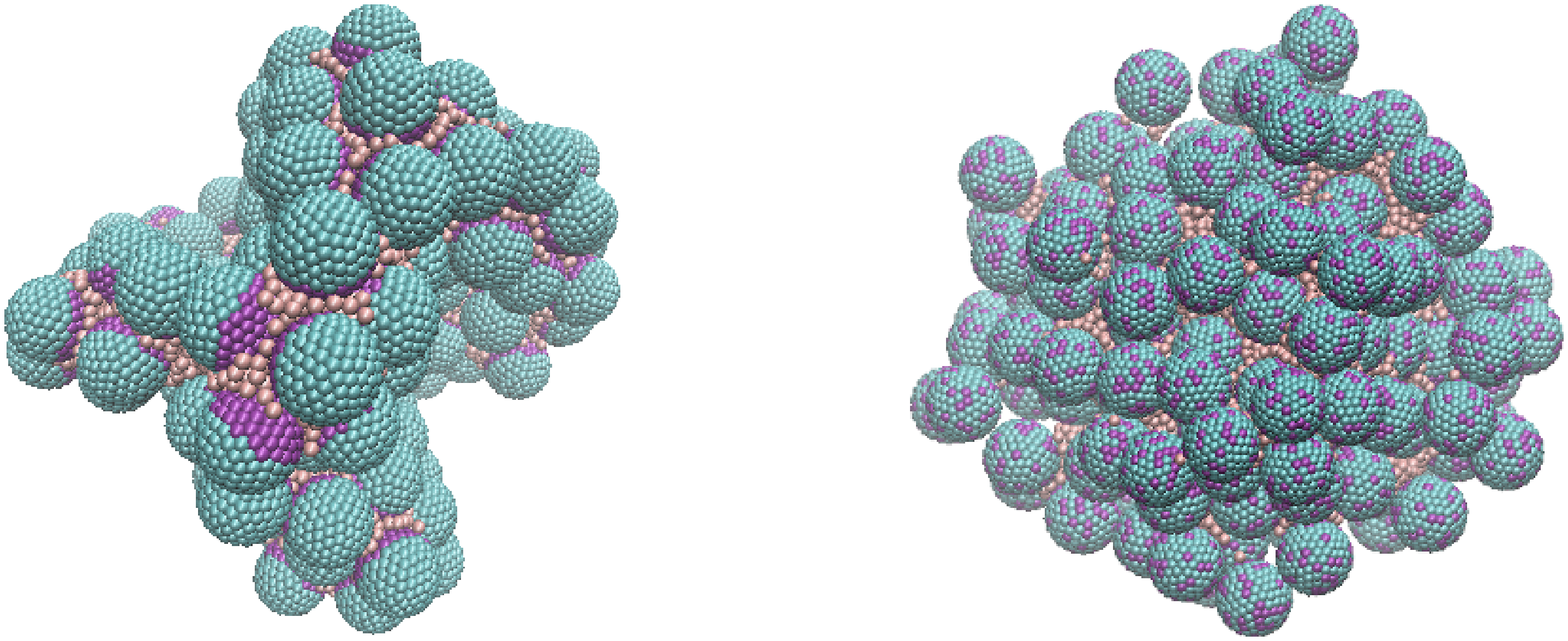}
\caption{Snapshots of the final configurations $E_{20}$ of the buckling processes of water droplets 
armored with 160 spherical Janus (left) and homogeneous (right). The proportion of water beads removed is 
around $10$ percent in each evaporation. The equilibration time after each evaporation is reduced to $2\times 10^3$ 
timesteps (instead of $10^5$ as in the main text).}
\label{figS3}
\end{figure}

In Fig.~\ref{figS3}, we show qualitatively, for comparison, the evolution of the droplet volume 
when the equilibration time after each water bead removal is reduced. The proportion of water beads 
removed remains in $10$ percent in each stage. 
When the droplet is coated with homogeneous NPs (right panel), we observe that the shape of the droplet 
remains spherical, with some NPs desorbed in the organic solvent. This evolution is representative 
of the \textit{passive} role played by the homogeneous NPs and
is in agreement with experiments~\cite{GarbinLangmuir2011}. 
When the droplet is coated with Janus NPs (left panel), we observe a significant 
curved-shape deformation of the droplet, along with the abscence of NP release. 
This evolution is representative of the \textit{active} role 
played by the Janus NPs. The morphology of the droplet becomes noticeably crumpled, with large dimples, and 
no transition to crater-like depression is observed.
This results are consistent with the \textit{surface model} numerical analysis from Ref.~\cite{QuillietEPJE2012}, 
where more than one dimple may nucleate if the evaporation is \textit{rapid}, leading to metastable 
multi-indented shapes. Experimentally, the term \textit{rapid} may correspond to kinetic barriers, 
which prevent thermally activated coalescence between adjacent dimples.
Our result highlights the central role played by 
the relaxation time of the system after each evaporation in the evolution of the interface geometry.
\end{document}